\documentstyle[aps,prd,preprint,tighten]{revtex} 
\begin{document} 
\draft
\preprint{imsc/98/05/22}
%\twocolumn[\hsize\textwidth\columnwidth\hsize\csname@twocolumnfalse\endcsname
\title{Duality transformation for 3+1 dimensional Yang-Mills theory}
\author{Pushan Majumdar \thanks{e-mail:pushan@imsc.ernet.in} \and
\addtocounter{footnote}{1}
H.S.Sharatchandra \thanks{e-mail:sharat@imsc.ernet.in}}
\address{Institute of Mathematical Sciences,C.I.T campus Taramani. 
Madras 600-113} 
\maketitle
\begin{abstract} 
Dual form of 3+1 dimensional Yang-Mills theory is obtained as another SO(3) 
gauge theory. Duality transformation is realized as a canonical 
transformation. The non-Abelian Gauss law implies the corresponding Gauss law 
for the dual theory. The dual theory is non-local. There is a non-local 
version of  Yang-Mills theory which is self dual.
\end{abstract}
%]
\section{Introduction}
Duality transformation plays an important role in many contexts in 
quantum field theory and statistical physics. It relates a model at a 
strong coupling or high temperature to another at weak coupling or low 
temperature. Therefore it provides a valuable tool in understanding some 
strongly interacting theories. In some cases, there is invariance under 
duality transformation. The standard example is the Ising model in 
two dimensions. In such a situation it provides further valuable 
information regarding properties of the system. Another mysterious 
aspect of duality transformations is that it often exposes topological 
degrees of freedom which play crucial roles in determining
properties of the system. A classic example is the 
Berezinsky-Kosterlitz-Thouless transition in two dimensional x-y model.

In this article we consider duality transformation of 3+1 dimensional 
Yang-Mills theory. Such transformations have already played crucial 
roles for understanding many aspects of gauge theories. Indeed the first 
examples of lattice gauge theories appeared as dual theories of certain 
Ising models \cite{Weg}.

Duality transformation is especially important for understanding the 
confinement aspects of gauge theories \cite{Kogut}. It is expected, and in some 
cases checked, that monopoles play a crucial role for this property. 
Three dimensional compact $U(1)$ gauge theory is a well understood 
example \cite{Pol1}. 

Duality transformation of an Abelian gauge theory gives the dual 
potential \cite{Banks} i.e. one which couples minimally to magnetic matter. 
Therefore it exposes the monopole degrees of freedom. This is brought 
out in a powerful way in four dimensional super symmetric gauge theories 
\cite{Witten}.

Deser and Teitelboim \cite{Deser} analyzed the possibility of duality invariance 
of 3+1 dimensional Yang-Mills theory in close analogy to Maxwell theory and 
concluded that invariance is not realized.
The first work to address duality transformation of 3+1 dimensional 
Yang-Mills theory retaining all the non-Abelian features
was by Halpern \cite{Hal}. Using complete axial gauge fixing, he brought 
out the crucial role played by the Bianchi identity. The dual theory was 
a gauge theory with a new gauge potential, though the action was non-local.

Another issue closely related to duality transformation is reformulation 
of the gauge theory dynamics using gauge invariant degrees of freedom. 
Several authors \cite{Sav} consider rewriting the functional integral using a 
gauge covariant second rank tensor. Anishetty, Cheluvraja, Sharatchandra and 
Mathur \cite{An} pointed out that $SO(3)$ 
lattice gauge theory in 2+1 dimensions is closely related to gravity. This 
can be used to formulate the dynamics using local gauge invariant 
degrees of freedom \cite{inv}. Similar situation is true in 3+1 dimensions 
also \cite{inv3}.

In this article we bring in new techniques which are useful 
for duality 
transformation of non-Abelian gauge theories. Though we use the language 
of functional integrals, our procedure can be stated directly for 
classical Yang-Mills theory. We adopt the Hamiltonian formalism. This 
is the most direct method for duality transformation in Maxwell's theory 
as reviewed in the first section. This also brings out the crucial role played 
by the Gauss law and the Hodge decomposition in duality transformation
which we developed in \cite{Hodge}.
 This approach automatically gives the dual theory as a 
SO(3) gauge theory, with a non-Abelian dual gauge field.

We also use generating functions
of canonical transformations to perform duality transformation (section III.B). 
We find that it is an extremely powerful technique for handling non-Abelian 
theories. It is very helpful for obtaining 
the implication of the non-Abelian Gauss law for the dual theory. It 
turns out that it is natural to treat the dual gauge field as a 
background gauge field of the Yang-Mills theory and vice-versa. (We use 
rescaled fields such that the gauge transformations do not involve the 
coupling constants.) Choosing the generating function to be invariant 
under a common gauge transformation, the Gauss law constraint simply 
goes over to a similar constraint in the dual theory (section III.C).
Another important issue is the gauge copy problem \cite{Wu,Copy}, i.e. gauge
inequivalent potentials which give the same non-Abelian magnetic field. 
In analogy to the Abelian case, we would like to replace $\vec{E}_{i}$, 
the non-Abelian electric field by $\vec{B}_{i}[C]$, the non-Abelian 
magnetic field of the dual gauge potential $C$. But if gauge copies are 
present, then this naive replacement runs into problems. 
We have argued in \cite{Copy} that there is only boundary degrees of freedom
 for the gauge field copies. As a consequence the number of 
degrees of freedom provided by $\vec{B}_{i}[C]$ are sufficient.

We explore the possibility of self duality of 3+1-dimensional Yang-Mills 
theory in section IV and conclude that it is absent. All the canonical 
transformations that we consider lead to a dual theory which is non-local.  

We summarize our results in sec V.

\section{Gauss law and duality transformation in Maxwell's theory}

Consider the free Maxwell theory. The extended phase space has
the canonical variables, the vector potential $A_i$ and the 
electric field $E_i$, $i=1,2,3$ with the Poisson bracket 
\begin{equation}
[A_i(x),E_j(y)]_{PB}= \delta_{ij}\delta(x-y).
\end{equation}
The Hamiltonian density is, 
\begin{equation}
H(x)=\frac{1}{2} (E_{i}^{2}(x)+B_i^{2}[A](x))
\end{equation}
where the magnetic field $B_i[A]=\epsilon_{ijk}\partial_j A_k$.
$A_i$ and $A_i+\partial_i \Lambda$ give rise to same $B_i[A]$.
The physical phase space is the subspace given by the 
Gauss law constraint, 
\begin{equation}
\partial_i E_i =0. 
\end{equation}

A very easy way of obtaining the dual theory is to solve the 
Gauss law constraint. We have the general solution,
\begin{equation}
E_i= \epsilon_{ijk}\partial_j C_k
\label{sqed}
\end{equation}
\noindent
We can compute the Poisson bracket of the new variable $C$ with 
the old variables as follows. We have the Poisson bracket 
\begin{equation}
[B_i(x),E_j(y)]_{PB}=- \epsilon_{ijk}\partial_k \delta(x-y).
\end{equation}
Substituting the above ansatz for $E$ we get
as a consistent solution the non-zero Poisson bracket
\begin{equation}
[B_i(x),C_j(y)]_{PB}= \delta_{ij} \delta(x-y).
\end{equation}
 Thus we have the new 
canonical pair $(C,\;{\cal E}=B[A])$  in contrast to the old set $(A,E)$.
In terms of this new pair the Hamiltonian takes the form
\begin{equation}
H(x)=\frac{1}{2} ({\cal E}^{2}_i(x)+B^{2}_i[C](x)).
\end{equation}
 Thus we have made a 
canonical transformation from the pair $(A,E)$ to $(C,B)$ and the 
Hamiltonian has the same form in terms the new variables. 
The analogy is complete since $C$ is also a gauge field (the dual 
gauge field), with $C_i(x)$ and $C_i(x)+\partial_i \lambda(x)$
giving rise to the same $B[C]$. This is the dual local gauge 
transformation. Also the new extended phase space has the dual 
Gauss law constraint 
\begin{equation}
\partial_i {\cal E}_i =0.
\end{equation} 
The old vector potential $A$ couples minimally to the electric 
currents. In contrast the new vector potential couples minimally 
to the magnetic current as can be verified by introducing sources.
Thus the dual symmetry is complete.

The duality transformation can be viewed as a canonical 
transformation induced by the generating function
\begin{equation}
S(A,C) \equiv \langle C|B[A]\rangle =\int \epsilon_{ijk} C_i \partial_j A_k
\end{equation}
of the old and the new coordinates $A$  and $C$ respectively.
We have the symmetry 
\begin{equation}
\langle C|B[A]\rangle =-\langle A|B[C]\rangle .
\end{equation}
 This 
is a very convenient technique for obtaining the new momentum and for 
computing the Poisson brackets of the old and the the new 
variables. We get the old and new momenta to be,
\begin{equation}
E_i=\frac{\delta S}{\delta A_i}= \epsilon_{ijk}\partial_j C_k =B_i[C],
\end{equation}
 and 
\begin{equation}
{\cal E}_i=-\frac{\delta S}{\delta C_i}=-B[A]_i
\end{equation}
respectively. The generating function is invariant under the old 
gauge transformation. This gives the identity, that for any $\lambda $
\begin{equation}
\int \partial_i \lambda \frac{\delta S}{\delta A_i}= 0.
\end{equation}
As $\lambda$ is arbitrary, it follows 
\begin{equation}
\partial_i \frac{\delta S}{\delta A_i}= 0,
\end{equation}
 which is the 
Gauss law constraint. This is a very convenient way of making the 
duality transformation preserving the Gauss law constraints. The 
generating function is also invariant under the new gauge 
transformation which implies the new Gauss law 
$ \partial_i {\cal E}_i =0$.
 
We extend and generalize these techniques for non-Abelian gauge theories.

\section{Techniques for duality transformation}
In this section we introduce various techniques useful for the duality 
transformation of non-Abelian gauge theories.
\subsection{Functional integral with phase space variables}
The Euclidean functional integral for 3+1-dimensional Yang-Mills theory is formally
\begin{equation}\label{Ef}
Z=\int\:{\cal D} 
A_{\mu}^{a}\;exp\{-\frac{1}{4g^{2}}\int\:\vec{F}_{\mu\nu}\cdot\vec{F}_{\mu\nu}\}
\end{equation}
where 
\begin{equation}\label{def}
\vec{F}_{\mu\nu}=\partial_{\mu}\vec{A}_{\nu}-\partial_{\nu}\vec{A}_{\mu}
+\vec{A}_{\mu}\times\vec{A}_{\nu}
\end{equation}
With this choice the gauge transformation does not involve the coupling 
constant. We could as well have started with the Minkowski space functional 
integral. However the Euclidean version makes the role of the 
non-Abelian Gauss law even more transparent.

Introducing an auxiliary field $E_{i}^{a}$, (\ref{Ef}) becomes
\begin{eqnarray}\label{aux}
Z&=&\int\:{\cal D} A_{0}^{a}{\cal D} A_{i}^{a} {\cal D} E_{i}^{a}\; exp
\int\{(\frac{-g^{2}}{2}\vec{E}_{i}\cdot\vec{E}_{i}-\frac{1}{2g^{2}}
\vec{B}_{i}[A]\cdot\vec{B}_{i}[A]) \nonumber \\
&&+i\vec{E}_{i}\cdot(\partial_{0}\vec{A}_{i}-D_{i}[A]\vec{A}_{0})\}
\end{eqnarray}
where 
\begin{equation}
D_{i}[A]=\partial_{i}+\vec{A}_{i}\times 
\end{equation}
 is 
the covariant derivative and 
\begin{equation}
\vec{B}_{i}[A]=\frac{1}{2}\epsilon_{ijk}( 
\partial_{j}\vec{A}_{k}-\partial_{k}\vec{A}_{j}+\vec{A}_{j}\times\vec{A}_{k})
\end{equation}
is the non-Abelian magnetic field.
Integration over $A_{0}$ gives
\begin{equation}\label{fgauss}
Z=\int\:{\cal D} A_{i}^{a}{\cal D} E_{i}^{a}\;\delta (D_{i}[A]E_{i})\;
 exp \{ \int (-{\cal H}+i\vec{E}_{i}\cdot\partial_{0}\vec{A}_{i})\}.
\end{equation}
Using the Feynman time slicing procedure, it is clear that $A_{i},E_{i}$ are the 
conjugate variables of the phase space and 
\begin{equation}\label{ham}
{\cal H}=\frac{1}{2} (g^{2}E^{2}+\frac{1}{g^{2}}B^{2})
\end{equation}
is the hamiltonian density. There are also three first class constraints, the 
non-Abelian Gauss law :
\begin{equation}\label{gauss1}
D_{i}[A]\vec{E}_{i}=0.
\end{equation}

\subsection{Duality transformation via a canonical transformation}
In close analogy to the Abelian case, we consider a change of variables 
from $E$ to ${\tilde C}$.
\begin{equation}\label{var}
\vec{E}_{i}=\epsilon_{ijk} D_{j}[A]\vec{\tilde C}_{k}
\end{equation}
where ${\tilde C}$ transforms homogeneously under gauge transformations. 
Naively ${\tilde C}_{i}^{a}$ is the canonical conjugate of the non-Abelian 
magnetic field $B_{i}^{a}$. This can be checked directly. Note that
\begin{equation}\label{pb}
[E_{m}^{d}(x), B_{i}^{a}(y)]_{PB}=\epsilon_{ijm}(\delta^{da}\partial_{j}
+\epsilon^{dab}A_{j}^{b})\delta(x-y).
\end{equation}
Using (\ref{var}), the left hand side is
\begin{equation}\label{pb2}
\epsilon_{ijm}(\delta^{de}\partial_{j}+\epsilon^{deb}A_{j}^{b})
[{\tilde C}_{m}^{e}(x), B_{i}^{a}(y)]_{PB}.
\end{equation}
This is consistent with 
\begin{equation}
[{\tilde C}_{m}^{e}(x), B_{i}^{a}(y)]_{PB}= 
\delta^{ea}\delta_{mi}\delta (x-y).
\end{equation}
An easy way to see this is by using the generator of canonical 
transformations 
\begin{equation}\label{can1}
S(A,{\tilde C})= \int\;{\tilde C}_{i}^{a}B_{i}^{a}[A]
\end{equation}
Then $E_{i}^{a}=\frac{\delta S}{\delta A_{i}^{a}}=\epsilon_{ijk} 
(D_{j}[A]C_{k})^{a}$ and the new momentum conjugate to the new variable 
${\tilde C}_{i}^{a}$ is 
\begin{equation}\label{con1}
{\cal E}_{i}^{a}=-\frac{\delta S}{\delta {\tilde C}_{i}^{a}}=-B_{i}^{a}[A].
\end{equation}

The great advantage of realizing duality transformation via a canonical 
transformation is that the phase space measure in the 
functional integral is invariant. 
\begin{equation}
{\cal D}A {\cal D}E={\cal D}C {\cal D}{\cal E}
\end{equation}
Also 
\begin{equation}
\sum p_{i}\dot{q}_{i}=\sum P_{i}\dot{Q}_{i} 
\end{equation}
and 
\begin{equation}
H^{\prime}(P,Q)=H(p(P,Q),q(P,Q))
\end{equation}
under a canonical transformation $(q,p)\rightarrow (Q,P)$. Therefore it is 
easy to express the 
exponent in equation (\ref{fgauss}) also in terms of the new variables.

\subsection{New Gauss law from the old Gauss law} 
In order to satisfy the Gauss law constraint (\ref{gauss1}), we need 
\begin{equation}\label{gsol}
\vec{B}_{i}[A]\times\vec{\tilde C}_{i}=0, 
\end{equation} 
where sum over $i$ is implied. Here we have used 
\begin{equation} 
\epsilon_{ijk}D_{j}[A]D_{k}[A]{\tilde
C}_i=\vec{B}_{i}[A]\times {\tilde C}_i. 
\end{equation} 
It is of interest to have the dual
field also a gauge field. With that in mind we introduce a new gauge field $C$. The
covariant derivative with respect to the gauge field $C$ can be written as 
\begin{equation}
D_{i}[C]=D_{i}[A]+(\vec{C}-\vec{A})_{i}\times.
\end{equation}
We also have the Bianchi identity
\begin{equation}\label{bianchi}
D_{i}[A]B_{i}[A]=0.
\end{equation} 
If we could preserve relation (\ref{con1}) in terms of the new field $C$, then we would get 
\begin{equation}\label{ng1}
D_{i}[C]\vec{\cal E}_{i}=-(\vec{C}-\vec{A})_{i}\times\vec{B}_{i}[A]
\end{equation}
(\ref{ng1}) together with (\ref{gsol}) 
immediately indicates that to get the new Gauss law, it is better to
rewrite ${\vec {\tilde C}}$ as ${\vec C}-{\vec A}$. This changes our ansatz (\ref{var}) to 
\begin{equation}\label{nans}
\vec{E}_{i}=\epsilon_{ijk}D_{j}[A](\vec{C}-\vec{A})_{k} .
\end{equation} 
This corresponds to the generating function 
\begin{equation}\label{gen1}
S(A,C)=\int\:(\vec{C}-\vec{A})_{i}\vec{B}_{i}[A] .
\end{equation} 
With this choice the old Gauss law (\ref{gauss1}) simply goes over to the new Gauss law
\begin{equation}\label{ngauss} 
D_{i}[C]\vec{\cal E}_{i}=0. 
\end{equation} 
Such a feature is very useful for the duality transformation. It can be easily realized in
general as shown below. 
In ansatz (\ref{var}), $C$ transforms homogeneously (as an isotriplet vector 
field) under the $A$-gauge transformation, whereas $A$ transforms inhomogeneously. 
\begin{equation}\label{gauge1} 
\delta A_{i}=D_{i}[A]\Lambda
\end{equation}
In contrast, in ansatz (\ref{nans})
$C$ transforms as a gauge field under $A$-gauge transformations. Note that 
if $C$ and $A$ both transform as gauge fields, $\alpha C + (1-\alpha)A$ 
also transforms like a gauge field for any choice of a real parameter 
$\alpha$. 
However $(C-A)$ transforms homogeneously, i.e. as a matter field in the adjoint
representation. Consider a 
canonical transformation $S(A,C)$ which is gauge invariant under the
common gauge transformations as in equation (\ref{gen1}). Some choices 
of terms in $S(A,C)$ are 
\begin{eqnarray}\label{terms}  
(a)&\epsilon_{ijk}(\vec{A}_{i}\cdot\partial_{j}\vec{A}_{k}+ \frac{1}{3} 
\vec{A}_{i}\cdot \vec{A}_{j}\times\vec{A}_{k})&\equiv {\cal CS}[A] \nonumber \\
(b)&\epsilon_{ijk}(\vec{C}_{i}\cdot\partial_{j}\vec{C}_{k}+ \frac{1}{3}
\vec{C}_{i}\cdot \vec{C}_{j}\times\vec{C}_{k})&\equiv {\cal CS}[C] \nonumber \\
(c)&(\vec{C}-\vec{A})_{i}\cdot\vec{B}_{i}[A]& \\
(d)&\epsilon_{ijk}\frac{1}{3!}(\vec{C}-\vec{A})_{i}\cdot 
(\vec{C}-\vec{A})_{j} \times (\vec{C}-\vec{A})_{k}&\equiv det(C-A).\nonumber
\end{eqnarray}
Here ${\cal CS}$ is the Chern-Simons density. Since 
\begin{equation}
\frac{\delta {\cal CS}[A]}{\delta A_{i}}=B_{i}[A],
\end{equation}
 it contributes a piece which is independent of $C$ to $E_{i}$.
Note that the functional integral (\ref{fgauss}) is insensitive to shifts
\begin{equation}
E_{i}\rightarrow E_{i}+\alpha B_{i}[A]
\end{equation}
 where $\alpha$ is an arbitrary real 
parameter. First of all, the Gauss law condition 
\begin{equation}
D_{i}[A]\vec{E}_{i}=0
\end{equation}
 does not change as a consequence of the Bianchi identity (\ref{bianchi}). 
Next, the 
term $E_{i}\dot{A}_{i}$ changes by 
\begin{equation}
\alpha B_{i}[A]\dot{A}_{i}=\alpha
\frac{\partial}{\partial t}{\cal CS}[A]. 
\end{equation}
This being a total derivative, does 
not matter. (This conclusion is not correct when instanton number \cite{Pol3} is 
non-zero.) This invariance is reflected in the possible addition of 
${\cal CS}[A]$ (\ref{terms} $a$) to the generating function $S[A,C]$

 Invariance of $S(A,C)$ under simultaneous gauge transformation of $A$ 
(\ref{gauge1}) and $C$, where, 
\begin{equation}\label{gauge2}
\delta \vec{C}_{i}=D_{i}[C]\vec{\Lambda}
\end{equation}
implies 
\begin{equation}\label{inv}
\int\:\left \{ ({\cal D}_{i}[A]\Lambda)^{a}\frac{\delta S}{\delta A_{i}^{a}}
+({\cal D}_{i}[C]\Lambda)^{a}\frac{\delta S}{\delta C_{i}^{a}}\right\}=0
\end{equation}
As this is true for any arbitrary choice of $\Lambda$, we get,
\begin{equation}\label{invgauss}
D_{i}[A]\vec{E}_{i}=D_{i}[C]\vec{\cal E}_{i}
\end{equation}
so that the old Gauss law constraint implies the new Gauss law constraint.
Another advantage of such a choice of $S(A,C)$ is that the dual field 
$C$ appears as a background gauge field for $A$ and vice-versa.

The new gauss law may be realized through an auxiliary field $C_{0}$ 
which would play the role played by $A_{0}$ in (\ref{aux}). This 
naturally leads to the action functional formulation of the dual theory, 
once we integrate over ${\cal E}_{i}$:
\begin{eqnarray}\label{daux}
Z&=&\int\:{\cal D}C_{0}{\cal D}C_{i}{\cal D}{\cal E}_{i}\;
exp\:\int \left \{ -H^{\prime}[C,{\cal E}]
+i(\partial_{0}\vec{C}-D_{i}[C]\vec{C}_{0})\cdot
\vec{\cal E}_{i} \right \} \nonumber \\
&=& \int\:{\cal D}C_{0}{\cal D}C_{i}\;exp\:(-S[C_{0},C_{i}])
\end{eqnarray}
where $S[C_{0},C_{i}]$ is gauge invariant under the full gauge 
transformation, $\delta \vec{C}_{\mu}=D_{\mu}[C]\vec{\Lambda}$.

\subsection{Degrees of freedom}
The constraint equation (\ref{gsol}) can be handled in a different way.
In the generic case where $det\,B \equiv 
|B|$, the determinant of the $3\times 3$ matrix $B_{i}^{a} (i,a=1,2,3)$ is 
non-zero, 
it is easy to solve this constraint on $C$ \cite{Hodge}. Use $B_{i}^{a}$ to ``lower"
the color index in $C_{i}^{a}$.
\begin{equation}\label{lower1}
C_{i}^{a}=C_{ij}B_{j}^{a}.
\end{equation}
Equation (\ref{gsol}) is satisfied if and only if $C_{ij}$ is a 
symmetric tensor. This corresponds to the choice
\begin{equation}\label{can2}
S(A,C)= \int\;C_{ij}b_{ij}
\end{equation}
where $C_{ij}$ would be the new coordinates and $b_{ij}=\vec{B}_{i}[A]\cdot 
\vec{B}_{j}[A]$, the new conjugate momenta.

Thus the ``physical" phase space of Yang-Mills 
theory may be described in terms of the conjugate pair $C_{ij}, b_{ij}$ 
which are gauge invariant symmetric second rank tensors. Each of these 
have six degrees of freedom at each $x$ which appears to match the 
required degrees of freedom. The situation could have been more involved
because of the Wu-Yang ambiguities \cite{Wu}. But as was analyzed in \cite{Copy}
this is not a generic phenomenon. The equation
\begin{equation}\label{drie}
\epsilon_{ijk}D_{j}[A]e_{k}=0 . 
\end{equation}
does not have a continuous family of solutions. Therefore we can write
\begin{equation}\label{var2}
\vec{E}_{i}=\epsilon_{ijk} D_{j}[A](\vec{C}_{k}-\vec{A}_{k})
\end{equation}
Alternately we can use the decomposition of the form \cite{Hodge}
\begin{equation}\label{decomp1}
\vec{E}_{i}=\vec{B}_{i}[C]
\end{equation}
This seems to be closest to the choice in the Abelian case which had 
duality invariance. Note that 
\begin{equation}\label{newmag}
\vec{B}_{i}[C]=\vec{B}_{i}[A]+\epsilon_{ijk}D_{j}[A](\vec{C}-\vec{A})_{k}
+\frac{1}{2}\epsilon_{ijk}(\vec{C}-\vec{A})_{j}\times (\vec{C}-\vec{A})_{k}
\end{equation}
which corresponds to an expansion of $B_{i}[C]$ about a ``background gauge 
field" $A$ with $(\vec{C}-\vec{A})$ as the quantum fluctuation.
 If $E_{i}$ satisfies the Gauss law (\ref{gauss1}), so does 
$E_{i}-B_{i}[A]$. Therefore the ansatz (\ref{nans}) and (\ref{decomp1}) 
essentially differ through the last term on the right hand side of 
(\ref{newmag}). This is obtained by including the term $det(C_A)$
(\ref{terms} d) in the generating functional of the canonical transformation.

The choice (\ref{decomp1}) is appealing for many reasons. We have,
\begin{equation}\label{mono}
\int\:\frac{1}{2} E^{2}_{i}=\int\left ( \frac{1}{2}B^{2}_{i}[C]\right ) 
\end{equation}
also
\begin{equation}\label{totder}
\frac{\delta S}{\delta \vec{A}_{i}}\partial_{0}\vec{A}_{i}+
\frac{\delta S}{\delta \vec{C}_{i}}\partial_{0}\vec{C}_{i}
=\partial_{0}S,
\end{equation}
a total derivative, so that,
\begin{equation}\label{rest}
\int\:\vec{E}_{i}\partial_{0}\vec{A}_{i}= \int\:\vec{\cal 
E}_{i}\cdot\partial_{0}\vec{C}_{i} 
\end{equation}
Therefore the exponent in (\ref{aux}) can be expressed easily in terms 
of the new variables as before.

\section{Duality Transformation}
In Maxwell theory we had 
duality invariance because $E_{i}=B_{i}[C]$ and ${\cal E}_{i}=-B_{i}[A]$. 
Such a simple interchange does not work for the non-Abelian case as seen 
from equations (\ref{con1}) and (\ref{nans}). Note that if we add ${\cal CS}[A]$, 
equation (\ref{terms}) to the generating function (\ref{gen1}), we can make
\begin{equation}\label{newE}
\vec{E}_{i}=\vec{B}_{i}[A] + \epsilon_{ijk}D_{j}[A](\vec{C}-\vec{A})_{k}.
\end{equation}
As seen from (\ref{newmag}) the quadratic term in $(C-A)$ is missing.

We now weaken our requirement. It is sufficient if,
\begin{equation}\label{symm}
g^{2}E^{2}+\frac{1}{g^{2}}B^{2}[A]=g^{2}B^{2}[C] +\frac{1}{g^{2}}{\cal E}^{2}
\end{equation}
If we use a generating function $S(A,C)$, we require
\begin{equation}\label{gensymm}
-g^{2}\left (\frac{\delta S}{\delta A_{i}}\right )^{2} +\frac{1}{g^{2}} 
\left (\frac{\delta S}{\delta C_{i}}\right )^{2}
=-g^{2}B^{2}[C] + \frac{1}{g^{2}}B^{2}[A].
\end{equation}
Consider the $g=1$ case.
Now equation (\ref{gensymm}) can be rewritten as
\begin{eqnarray}\label{regensymm}
\frac{\delta S}{\delta \left ( \frac{A+C}{2}\right )_{i}}
\frac{\delta S}{\delta \left ( \frac{A-C}{2}\right )_{i}}&=&
\epsilon_{ijk}D_{j}\left [ \frac{A+C}{2}\right]
\left ( \frac{\vec{A}-\vec{C}}{2}\right )_{k}\cdot
\left \{\vec{B}_{i}\left[\frac{A+C}{2}\right] \right . \nonumber \\
&&\left . +\frac{1}{2} \epsilon_{ijk}\left ( \frac{\vec{A}-\vec{C}}{2}
\right)_{j}\times \left ( \frac{\vec{A}-\vec{C}}{2}\right )_{k}\right \}
\end{eqnarray}
using equation (\ref{newmag}) for the background gauge field 
$(\frac{C+A}{2})$. It is amusing to note that the generating function
\begin{equation}\label{false}
S\left(\frac{A+C}{2},\frac{A-C}{2}\right)=\left(\frac{\vec{A}-\vec{C}}{2} 
\right)_{i} \cdot
\vec{B}_{i}\left[\frac{A+C}{2}\right] + det \left(\frac{A-C}{2}\right)
\end{equation}
gives the right hand side of the above equation, but with the opposite sign.
Self duality is achieved in the Abelian case by using
\begin{equation}\label{abelian}
S={\cal CS}\left( \frac{C+A}{2}\right) - {\cal CS}\left(\frac{C-A}{2}\right).
\end{equation}
The non-Abelian case should have something similar and not (\ref{false}). 
Unfortunately there is no $S$ satisfying (\ref{regensymm}). As a consequence 
self duality is ruled out.

We consider generating functions
\begin{eqnarray}\label{fgen}
S(A,C)&=&\alpha_{1}{\cal CS}(A) +\alpha_{2}{\cal CS}(C) +\alpha_{3} 
(\vec{A}-\vec{C})_{i}\cdot \vec{B}_{i}[A] \nonumber \\ 
&&+\frac{\alpha_{4}}{2}\epsilon_{ijk}(\vec{A}-\vec{C})_{i}\cdot 
D_{j}[A](\vec{A}-\vec{C})_{k} +\alpha_{5} det(A-C). 
\end{eqnarray}
where $\alpha_{1},\ldots \alpha_{5}$ are arbitrary real parameters for the present.
Now we get
\begin{eqnarray}\label{fE}
\vec{E}_{i}&=&\beta_{1}\vec{B}_{i}[A] +\beta_{2}\epsilon_{ijk} D_{j}[A]
(\vec{A}-\vec{C})_{k} \nonumber \\
&&\hspace*{15 mm} + \frac{\beta_{3}}{2}\epsilon_{ijk} 
(\vec{A}- \vec{C})_{j}\times (\vec{A}-\vec{C})_{k} \\
\vec{\cal E}_{i}&=&\gamma_{1}\vec{B}_{i}[A]+\gamma_{2}\epsilon_{ijk} D_{j}[A] 
(\vec{A}-\vec{C})_{k} \nonumber \\
&&\hspace*{15 mm} + \frac{\gamma_{3}}{2}\epsilon_{ijk}(\vec{A}-\vec{C})_{j}
\times (\vec{A}-\vec{C})_{k}
\end{eqnarray}
where $\beta_{1}=\alpha_{1}+\alpha_{3};\; \beta_{2}=\alpha_{3}+\alpha_{4};
\; \beta_{3}=\alpha_{4}+\alpha_{5};\;$ and $\gamma_{1}=-\alpha_{2}+\alpha_{3};
\; \gamma_{2}=\alpha_2+\alpha_{4};\; \gamma_{3}=-\alpha_2+\alpha_{5}$.
For no choice of the parameters $\alpha_{1},\ldots \alpha_{5}$ do we get a local 
Hamiltonian in the dual variables. We illustrate this for a specific choice,
$\alpha_{2}=\alpha_{4}=\alpha_{5}=0,\:\alpha_1=-2$ and $\alpha_{3}=1$. We 
get $\vec{\cal E}_{i}=\vec{B}_{i}[A]$ but $\vec{E}_{i}=-\vec{B}_{i}[C] 
-\frac{1}{2}\epsilon_{ijk}
(\vec{A}-\vec{C})_{j}\times (\vec{A}-\vec{C})_{k}$. Therefore the dual 
action becomes
\begin{equation}\label{faction}
g^{2}\{B_{i}[C]+\frac{1}{2}\epsilon_{ijk}(\vec{A}-\vec{C})_{j}\times 
(\vec{A}-\vec{C})_{k}\}^{2} +\frac{1}{g^{2}}{\cal E}^{2}.
\end{equation}
$(A-C)$ may be regarded as a non-local functional of the dual variables 
$(C,{\cal E})$; solution of
\begin{equation}\label{nonlocal}
\epsilon_{ijk} D_{j}[C] 
(\vec{A}-\vec{C})_{k} +\frac{1}{2}\epsilon_{ijk}(\vec{A}-\vec{C})_{j}\times 
(\vec{A}-\vec{C})_{k}=\vec{\cal E}_{i} - \vec{B}_{i}[C]
\end{equation}

Consider a modified Yang-Mills Hamiltonian
\begin{equation}\label{modham}
H=\int\left(\frac{1}{2}g^{2}\vec{E}_{i}^{2}+\frac{1}{2g^{2}}\vec{\cal E}_{i}^{2}
\right) 
\end{equation}
where it is presumed that ${\cal E}_{i}=-\frac{\delta S}{\delta C_{i}}$ is 
expressed in terms of $(A,E)$. This theory would be self dual, 
if the generating function $S(A,C)$ is symmetric 
under the interchange $A\leftrightarrow C$. A simple way of realizing this 
is to have $S$ (regarded as a functional of $(A+C)$ and $(A-C))$, even in 
$(A-C)$. For all choices of $S$ we have considered, the theory is non-local.

\section{Conclusion}

In this article we have constructed a dual form of the 3+1 Yang-Mills 
theory. We have argued that the functional integral using phase space 
variables is best suited for the purpose. Now the duality transformation 
can be realized as a canonical transformation. This provides a powerful tool, 
because the action and the measure in the dual variables as also the 
implications of the Gauss law constraint for the dual theory are easily written.
The dual theory is also a SO(3) gauge theory. 
The dual theory, though a SO(3) gauge 
theory, is a non-local 
theory. However Yang-Mills theory with a non-local action is self dual.
Our techniques for obtaining the dual theory may provide a firm basis for the 
computations of the confining properties in the dual QCD approach of Baker, Ball 
and Zachariasen \cite{Dual}.


\begin{references}
\bibitem{Weg} F. Wegner, J. Math. Phys. {\bf 12} (1971), 2259.
\bibitem{Kogut} John B. Kogut, Rev. Mod. Phys. {\bf 51} (1979), 659.
\bibitem{Pol1} A.Polyakov, Phys. Lett. {\bf B59} (1975), 82;
A.Polyakov, Nucl. Phys. {\bf B120} (1977), 429
\bibitem{Banks} T. Banks, R. Myerson and J. Kogut, Nucl. Phys. {\bf B129}
 (1977), 493;  J. L. Cardy, Nucl. Phys. {\bf B205} (1982), 1;
Manu Mathur and H. S. Sharatchandra, Phys. Rev. Lett. {\bf 66} (1991), 3097.
\bibitem{Witten} W. Seiberg and E. Witten, Nucl. Phys. {\bf B426} (1994),
19.
\bibitem{Deser} S. Deser and C. Teitelboim, Phys. Rev. {\bf D13} (1976), 1592.
\bibitem{Hal} M. B. Halpern, Phys. Rev. {\bf D16} (1977), 1798;
M. B. Halpern, Phys. Rev. {\bf D19} (1979), 517.
\bibitem{Sav} Y. Kazama and R. Savit, Phys. Rev. {\bf D21} (1980), 2916;
G. Chamlers and W. Siegel, hep-th/9712191.
\bibitem{An} 
R. Anishetty and H. S. Sharatchandra, Phys. Rev. Lett. {\bf 65} (1990), 813;
R. Anishetty, S. Cheluvraja, H. S. Sharatchandra and M. Mathur,
Phys. Lett. {\bf B314} (1993), 387.
\bibitem{inv} 
B. Gnanapragasam and H. S. Sharatchandra, Phys. Rev. {\bf D45} (1992), R1010;
R. Anishetty, Pushan Majumdar and H. S. Sharatchandra, Phys.Lett. {\bf B478} 
(2000), 373.
\bibitem{inv3}
R. Anishetty, Phys. Rev. {\bf D44} (1991), 1895;
F. A. Lunev, Phys. Lett. {\bf B295} (1992), 99;
F. A. Lunev, Mod. Phys. Lett. {\bf A9} (1994), 2281;
F. A. Lunev, J. Math. Phys. {\bf 37} (1996), 5351;
M. Bauer, D. Z. Freedman and P. E. Haagensen, Nucl. Phys. {\bf B428} 
(1994), 147;
P. E. Haagensen and K. Johnson, Nucl. Phys. {\bf B439} (1995), 597.
\bibitem{Hodge} Pushan Majumdar and H. S. Sharatchandra, 
Phys.Rev. {\bf D58} (1998), 067702.
\bibitem{NR} E. T. Newman and C. Rovelli, Phys. Rev. Lett.
{\bf 69} (1992), 1300.
\bibitem{Wu} T. T. Wu and C. N. Yang, Phys. Rev. {\bf D12} (1975), 3845.
\bibitem{FK} D. Z. Freedman and R. R. Khuri, Phys. Lett. {\bf B329} (1994), 
263.
\bibitem{Copy} Pushan Majumdar and H. S. Sharatchandra, imsc/98/04/14, 
hep-th/9804091 (1998).
\bibitem{Pol3} A. A. Belavin, A. M. Polyakov, A. S. Shvarts, Yu. S. Tyupkin 
Phys. Lett. {\bf 59B} (1975), 85.
\bibitem{Dual} M. Baker, J. S. Ball and F. Zachariasen, Phys. Rep. {\bf 209} 
(1991), 73.
\end{references}
\end{document}